\documentclass[12pt,thmsa]{article}

\begin{document}

\title{Fractional spin through quatum (super)Virasoro algebras.}
\author{M. Mansour \thanks{%
mansour70@mailcity.com} and E. H. Zakkari \thanks{%
hzakkari@yahoo.fr} \\
Laboratory of Theoretical Physics\\
University Mohamed V\\
PO BOX 1014\\
Rabat, Morocco.}
\date{}
\maketitle

\begin{abstract}
The splitting of a $Q$-deformed boson, in the $Q\to q=e^{\frac{\mathrm{2\pi i%
}}{\mathrm{k}}}$ limit, is discussed. The equivalence between a $Q$-fermion
and an ordinary one is established. The properties of the quantum
(super)Virasoro algebras when their deformation parameter $Q$ goes to a root
of unity, are investigated. These properties are shown to be related to
fractional supersymmetry and $k$-fermionic spin.
\end{abstract}

\newpage\

\section{Introduction}

Recently, quantum groups and quantum algebras $[1,2]$ have played a central
role in the development of various arena of mathematics and theoretical
physics. Indeed, this new mathematical structure emerge in several context:
scattering method $[3]$ as well as Yang-Baxter equations in rational
conformal field theory $[4,5,6,7]$. The representation theory of quantum
(super)algebras has been also an object of intensive studies. A vailable are
the results for the oscillator representation of quantum (super) algebras.
The latters are obtained through consistent realization involving deformed
Bose and Fermi operators $[8,9].$

In connection with quantum group theory, a new interpretation of fractional
supersymetry has been developed in references $[10,11,12,13,14]$. In these
works, the authors proved that the one-dimensional superspace is isomorphic
to the braided line when the deformation parameter is a root of unity. In
this case the fractional supersymetry is identified as translational
invariance along this line. It is remarkable that at the limit $q=e^{\frac{%
2i\pi }k}$, the braided line $[15,16,17]$ is generated by a generalized
Grassman variable and an ordinary even one.

Following the same technique, it is proved that internal spin arises
naturally at a certain limit of the finite the Q-deformed algebra $%
U_Q(sl(2)) $ $[18].$ In deed, using Q-Schwinger realization, it is shown
that the $U_Q(sl(2))$ is equivalent to a direct product of the finite
undeformed algebra $U(sl(2))$ and the deformed one $U_q(sl(2))$ (note that $%
U_Q(sl(2))$ $=$ $U_q(sl(2))$ at $Q=q).$ Since there exist $Q$-oscillator
realization of all deformed enveloping algebras $U_Q(g),$ it is reasonable
to expect these to admit analogous decompositions (or splitting) when $%
Q\rightarrow q.$

In recent years, much interest has been made in the study of the
infinite dimensional algebras. The latters appeared in string
theories and the two dimensional conformal field theory. These
symmetries have been shown to be related to the Virasoro algebra.
the two dimensional superconformal field theory can be treated
from a group-theoretical point of view, the basic ingredient is
the two-dimensional superconformal algebra which is infinite
dimensional and called the superVirasoro algebra, see $[19]$ and
the references therin. The central extension of superVirasoro
leads to so called Ramond-Neveu-Schwarz algebras $[20,21].$ A
supersymmetric extension of the work of Curthright and Zachos is
given in $[22]$ by using a quantum superspace approach $[23,24].$

In the same spirit of quantum (affine) algebras $[25,26]$, it is more
interesting to study the splitting of deformed (super)Virasoro algerbas at a
root of unity in order to describe the (super)conformal symmetries at this
limit, this is the aim of our present paper.

In this work, we investigate the property of splitting for the deformed
(super)Virasoro algebras in the $Q\rightarrow q$ limit . At the first steps
we start in section $(2)$ by defining the k-fermionic algbras. In section $%
(3)$ we recall some preliminaries results concerning the property of $Q$%
-boson decomposition in the $Q\rightarrow q$ limit. We shall first
discuss the way in which one obtains two independent objects (an
ordinary boson and a $k$-fermion) from one Q-deformed boson when
$Q$ goes to a root of unity. We also show the equivalence between
a $Q$- deformed fermion and a conventional (ordinary or
undeformed) fermion. Using the $Q$-Schwinger realization which
presents an interesting properties in the $Q\rightarrow q$ limit,
we have analyzed the decomposition for the $Q$-deformed Virasoro
algebra in section $(4)$, and the quantum superVirasoro algebra in section $%
(5).$

\section{Preliminaries about k-fermionic algebra.}

The $q$-deformed bosonic (k-fermion) algebra $\Sigma _q$ generated by $A^{+}$%
, $A^{-}$ and number operator $N$ is given by:

\begin{equation}
A^{-}A^{+}-qA^{+}A^{-}=q^{-N}
\end{equation}

\begin{equation}
A^{-}A^{+}-q^{-1}A^{+}A^{-}=q^N
\end{equation}

\begin{equation}
q^NA^{\pm }q^{-N}=q^{\pm 1}A^{\pm }
\end{equation}

\begin{equation}
q^Nq^{-N}=q^{-N}q^N=1,
\end{equation}
where the deformation parameter:

\begin{equation}
q=e^{\frac{2i\pi }l},\; l\in N-\{0,1\},
\end{equation}

is a root of unity.

The annihilation operator $A^{-}$ is hermitian conjugate to creation
operator $A^{+}$ and $N$ is hermitian also. From equations $(1)-(4)$, it is
easy to have the following relations:

\begin{equation}
A^{-}(A^{+})^n=[[n]]q^{-N}(A^{+})^{n-1}+q^n(A^{+})^nA^{-}
\end{equation}

\begin{equation}
(A^{-})^nA^{+}=[[n]](A^{-})^{n-1}q^{-N}+q^nA^{+}(A^{-})^n,
\end{equation}
where the notation $[[$ $]]$ is defined by:

\begin{equation}
\lbrack [n]]=\frac{1-q^{2n}}{1-q^2}
\end{equation}

We introduce a new variable $k$ defined by:
\begin{equation}
k=l\;\; \hbox{for  odd  values  of}\;\; l,
\end{equation}

\begin{equation}
k\,=\,\frac l2\;\;\hbox{for even values of}\;\;l,
\end{equation}
such that for odd $l$ (resp. even $l$ ), we have $q^k=1$ $($resp. $q^k=-1).$
In the particular case $n=k$, equations $(6)-(7)$ permit us to have:

\begin{equation}
A^{-}(A^{+})^k=\pm (A^{+})^kA^{-}
\end{equation}

\begin{equation}
(A^{-})^kA^{+}=\pm A^{+}(A^{-})^k,
\end{equation}
and the equations $(1)-(5)$ yield to:

\begin{equation}
q^N(A^{+})^k=(A^{+})^kq^N
\end{equation}

\begin{equation}
q^N(A^{-})^k=(A^{-})^kq^N
\end{equation}
One can show that the elements $(A^{-})^k$ and $(A^{+})^k$ are the elements
of the centre of $\sum_q$ algebra (odd values for $l$); and the irreducible
representations are $k$-dimensional. These two properties lead to:
\begin{equation}
(A^{+})^k=\alpha I
\end{equation}

\begin{equation}
(A^{-})^k=\beta I.
\end{equation}

The extra possibilities parameterized by:

\[
(1) \;\; \alpha \, =\, 0, \;\; \beta \neq 0
\]

\[
(2) \;\; \alpha \neq 0, \;\; \beta =0
\]

\[
(3)\;\;\alpha \neq 0,\;\;\beta \neq 0,
\]
are not relevant for the considerations of this paper. In the two cases $(1)$
and $(2)$ we have the so-called semi-periodic (semi-cyclic) representation
and the case $(3)$ correspond to the periodic one. In what follows, we are
interested in a representation of the algebra $\sum_q$ such that the
following:

\[
(A^{\mp })^k=0,
\]
is satisfied. We note that the algebra $\sum_{-1}$ obtained for $k=2$,
correspond to ordinary fermion operators with $(A^{+})^2=0$ and $(A^{-})^2=0$
which reflects the exclusion's Pauli principle. In the limit case where $%
k\rightarrow \infty $, the algebra $\sum_1$ correspond to the ordinary
bosons. For other values of $k$, the $k$-fermions operators interpolate
between fermions and bosons, these are also called anyons with fractional
spin in the sense of Majid $[15,16,17]$.

\section{Fractional spin through Q-boson.}

In the previous section, we have worked with $q$ at root of unity. In this
case, quantum oscillator $(k$-fermionic$)$ algebra exhibit a rich
representation with very special properties different from the case where $q$
is generic. So, in the first case the Hilbert space is finite dimensional.
In contrast, where $q$ is generic, the Fock space is infinite dimensional.
In order to investigate the decomposition of $Q$-deformed boson in the limit
$Q\rightarrow e\frac{2i\pi }k$ we start by recalling the $Q$-deformed
algebra $\Delta _Q$.

The algebra $\Delta _Q$ generated by an annihilation operator $B^{-}$, a
creation operator $B^{+}$ and a number operator $N_B$:

\begin{equation}
B^{-}B^{+}-QB^{+}B^{-}=Q^{-N_B}
\end{equation}

\begin{equation}
B^{-}B^{+}-Q^{-1}B^{+}B^{-}=Q^{N_B}
\end{equation}

\begin{equation}
Q^{N_B}B^{+}Q^{-N_B}=QB^{+}
\end{equation}

\begin{equation}
Q^{N_B}B^{-}Q^{-N_B}=Q^{-1}B^{-}
\end{equation}

\begin{equation}
Q^{N_B}Q^{-N_B}=Q^{-N_B}Q^{^{+}N_B}=1.
\end{equation}

From the above equations, we obtain:
\begin{equation}
\lbrack
Q^{-N_B}B^{-},[Q^{-N_B}B^{-},[....[Q^{-N_B}B^{-},(B^{+})^k]_{Q^{2k}}...]_{Q^4}]_{Q^2}]=Q^{%
\frac{k(k-1)}2}[k]!
\end{equation}
where the $Q$-deformed factorial is given by:

\begin{equation}
\lbrack k]!=[k][k-1][k-2]...............[1],
\end{equation}
and:

\[
\lbrack 0]!=1
\]

\[
\lbrack k]=\frac{Q^k-Q^{-k}}{Q-Q^{-1}}\hbox{ .}
\]

The $Q$-commutator, in equation $(22)$, of two operators $A$ and $B$ is
defined by:

\[
\lbrack A,B]_Q=AB-QBA
\]

The aim of this section is to determine the limit of $\Delta _Q$ algebra
when $Q$ goes to the root of unity $q$. The starting point is the limit $%
Q\rightarrow q$ of the equation $(22),$

\[
\lim_{Q\rightarrow q}\frac
1kQ^{-N_B}[Q^{-N_B}B^{-},[Q^{-N_B}B^{-},[....[Q^{-N_B}B^{-},(B^{+})^k]_{Q^{2k}}...]_{Q^4}]_{Q^2}]
\]

\begin{equation}
=\lim_{Q\rightarrow q}\frac{Q^{\frac{k(k-1)}2}}{[k]!}[%
Q^{-N_B}(B^{-})^k,(B^{+})^k]=q^{\frac{k(k-1)}2}
\end{equation}

This equation can be reduced to:

\begin{equation}
\lim_{Q\rightarrow q}[\frac{Q^{\frac{kN_B}2}(B^{-})^k}{([k]!)^{\frac 12}},%
\frac{(B^{+})^kQ^{\frac{kN_B}2}}{([k]!)^{\frac 12}}]=1.
\end{equation}

Since $q$ is a root of unity, it is possible to change the sign on the
exponent of $q^{\frac{kN_B}2}$ terms in the above equation.

We define the operators as in $[18]$:

\begin{equation}
b^{-}=\lim_{Q\rightarrow q}\frac{Q^{\pm \frac{kN_B}2}}{([k]!)^{\frac 12}}%
(B^{-})^k,\,b^{+}=\lim_{Q\rightarrow q}\frac{(B^{+})^kQ^{^{\pm }\frac{kN_B}2}%
}{([k]!)^{\frac 12}},
\end{equation}
which lead to an ordinary boson algebra noted $\Delta _0$, generated by:

\begin{equation}
\lbrack b^{-},b^{+}]=1.
\end{equation}

The number operator of this new bosonic algebra defined as the usual case, $%
N_b=b^{+}b^{-}$. At this stage we are in a position to discuss the splitting
of $Q$-deformed boson in the limit $Q\rightarrow q$ . Let us introduce the
new set of generators given by:

\begin{equation}
A^{-}=B^{-}q^{-\frac{kN_b}2}
\end{equation}

\begin{equation}
A^{+}=B^{+}q^{-\frac{kN_b}2}
\end{equation}

\begin{equation}
N_A=N_B-kN_b,
\end{equation}
which define a $k$-fermionic algebra:

\begin{equation}
\lbrack A^{+},A^{-}]_{q^{-1}}=q^{N_A}
\end{equation}

\begin{equation}
\lbrack A^{-},A^{+}]_q=q^{-N_A}
\end{equation}

\begin{equation}
\lbrack N_A,A^{\pm }]=\pm A^{\pm }.
\end{equation}
It is easy to verify that the two algebras generated by the set of operators
$\{b^{+},b^{-},N_b\}$ and $\{A^{+},A^{-},N_A\}$ are mutually commutative. We
conclude that in the limit $Q\rightarrow q$ , the $Q$-deformed bosonic
algebra oscillator decomposes into two independent oscillators, an ordinary
boson and $k$-fermion; formally one can write:

\[
\lim_{Q\rightarrow q}\Delta _Q\equiv \Delta _0\otimes \Sigma _q,
\]
where $\Delta _0$ is the classical bosonic algebra generated by the
operators $\{b^{+},b^{-},$ $N_b\}.$

Similarly, we want to study the $Q$-fermion algebra at root of unity. To do
this, we start by considering the $Q$- deformed fermionic algebra, noted $%
\Xi _Q$:

\begin{equation}
F^{-}F^{+}+QF^{+}F^{-}=Q^{N_{F}}
\end{equation}

\begin{equation}
F^{-}F^{+}+Q^{-1}F^{+}F^{-}=Q^{-N_F}
\end{equation}

\begin{equation}
Q^{N_F}F^{+}Q^{-N_F}=QF^{+}
\end{equation}

\begin{equation}
Q^{N_F}F^{-}Q^{-N_F}=Q^{-1}F^{-}
\end{equation}

\begin{equation}
Q^{N_F}Q^{-N_F}=Q^{-N_F}Q^{N_F}=1
\end{equation}

\begin{equation}
(F^{+})^2=0, \; \; (F^{-})^2=0
\end{equation}

We define the new fermionic operators as follow:

\begin{equation}
f^{+}=\lim_{Q\rightarrow q}F^{+}Q^{^{-}\frac{N_F}2}
\end{equation}

\begin{equation}
f^{-}=\lim_{Q\rightarrow q}Q^{^{-}\frac{N_F}2}F^{-}.
\end{equation}

By a direct calculus, we obtain the following anti-commutation relation:

\begin{equation}
\{f^{-},\, f^{+}\}=1.
\end{equation}

Moreover, we have the nilpotency condition:

\begin{equation}
(f^{-})^2=0, \;\;(f^{+})^2=0.
\end{equation}

Thus, we see that the $Q$-deformed fermion reproduce the conventional
(ordinary) fermion. The same convention notation permits us to write:

\[
\lim_{Q\rightarrow q}\Xi _Q\equiv \Sigma _{-1}
\]

\section{The deformed centerless Virasoro algebra}

We apply now the above results to derive the property of decomposition of
the quantum Virasoro algebra in the $Q\rightarrow q$ limit. Recalling that
the classical Virasoro algebra $(vir)$ is generated by the following set of
generators $\{l_n,n\in Z\}$ such that:
\begin{equation}
\lbrack l_n,l_m]=(m-n)l_{n+m}.  \label{e47}
\end{equation}

It is well known that the algebra $(44)$ can be realized by considering the
Schwinger construction. This realization involve one classical (undeformed)
bosonic algebra $\{b^{+},b^{-},N_b\}$ as follows:
\begin{equation}
l_n=(b^{+})^{n+1}b^{-},~~~~~~~~n\ge -1.  \label{e48}
\end{equation}

Recently, a lot of attention has been paid to the $Q$-deformation of the
centreless Virasoro algebra $[27,28,29,30]$ and its central extension $%
[30,31,32]$. Recalling that the one parameter deformation of the centerless
Virasoro algebra $(vir_Q)$ is given by:
\begin{equation}
\lbrack L_n,L_m]_{(Q^{m-n},Q^{n-m})}=[m-n]L_{n+m},  \label{e49}
\end{equation}
where,
\[
\lbrack A,B]_{(\alpha ,\beta )}=\alpha AB-\beta BA,
\]

\[
\lbrack x]=\frac{q^x-q^{-x}}{q-q^{-1}}.
\]

A possible realization of $Q$-Virasoro generators is given by:
\begin{equation}
L_n=Q^{-N}(B^{+})^{n+1}B^{-},  \label{e50}
\end{equation}
where $B^{+}$ and $B^{-}$ are $Q$-deformed creation and annihilation
operators respectively, generating the $Q$-bosonic algebra.

At this stage, our aim is to investigate the limit $Q\to q$ of the $Q$%
-deformed Virasoro algebra $vir_Q.$ As it is already mentioned in
the introduction, our analysis is based on the oscillator
representation. In the $Q\to q$ limit, the splitting of Q-deformed
boson $\{B^{+},B^{-},$ $N_B\}$
lead to a classical boson $\{b^{+},b^{-},N_b\}$ given by the equations $%
(26,27)$ and a $k$-fermion algebra $\{A^{+},A^{-},N_A\}$ given by $%
eqs(28-30).$ From the undeformed boson, we define the generators $j_n$:

\begin{equation}
j_n=(b^{+})^{n+1}b^{-},~~~~~~~~n\ge -1,
\end{equation}
which generates the classical Virasoro algebra:

\begin{equation}
\lbrack j_n,j_m]=(m-n)j_{n+m}
\end{equation}

From the remaining operators $\{A^{+},A^{-},N_A\}$, one can realize the
q-deformed Virasoro algebra $vir_q$:

\begin{equation}
\lbrack J_n,J_m]_{(q^{m-n},q^{n-m})}=[m-n]J_{n+m}
\end{equation}

Indeed, the generators defined by:

\begin{equation}
J_n=q^{-N_A}(A^{+})^{n+1}A^{-},
\end{equation}
generate the $vir_q$ algebra which is the same version of $vir_Q$ obtained
by simply setting $Q=q$, rather than by taking the limit as above.

The generators of classical Virasoro and $q$-deformed Virasoro
algebra are mutually commutative:

\begin{equation}
\lbrack J_k,j_l]=0.
\end{equation}

So, we obtain the following decomposition:

\[
\lim_{Q\rightarrow q}vir_Q=vir_q\otimes vir.
\]

We remark that the $q$-deformed Virasoro algebra exhibit some interesting
properties. Indeed, when $m-n=rl$ for any $r\in Z$ the equation $(50)$ is
reduced to:
\begin{equation}
\lbrack J_n,J_m]=0.  \label{e55}
\end{equation}

Noticing that, due to the nilpotency condition $(A^{+})^l=(A^{-})^l=0$, the
generators $J_n$ vanishes for any $n\ge l-1$. This fact constitutes an
interesting property of the deformed Virasoro algebra when the deformation
parameter is a root of unity. Namely, for particular value $l=3$, the $q$%
-deformed Virasoro algebra reduces to its subalgebra $su_q(2)$ generated by $%
\{J_0,J_1,J_{-1}\}$.

\section{The Quantum superVirasoro algebra}

The classical superVirasoro algebra is generated by the following set of
generators $\{L_n,G_n,F_n;n\in Z\}$ satisfying the defining relations $[19]$%
:
\begin{equation}
\begin{array}{c}
\lbrack L_n,L_m]=(n-m)L_{n+m} \\
~~~~[F_m,G_n]=G_{n+m} \\
\lbrack L_n,F_m]=-mF_{n+m} \\
~~[F_m,F_n]=0 \\
\lbrack L_m,G_n]=(m-n)G_{n+m} \\
~~~~[G_m,G_n]=0,
\end{array}
\label{e1}
\end{equation}

The $Z_2$ -grading on this superalgebra is defined by requiring that $%
deg(L_i)=$ deg$(F_i)=0$ and deg$(G_i)=1$; further, the bracket $[,]$ in
relations $(54)$ stands for a graded one:
\[
\lbrack x,y]=xy-(-1)^{deg(x)deg(y)}yx.
\]

The classical superalgebra $(54)$ can be realized by considering the
Schwinger construction.

This realization involve two undeformed algebras bosonic $%
\{b^{-},b^{+},N_b\} $ and fermionic $\{f^{-},f^{+},N_f\}$ one:

\begin{equation}
\begin{array}{c}
L_n=-(b^{+})^{n+1}b^{-} \\
G_n=(b^{+})^{n+1}f^{+}b^{-} \\
F_n=(b^{+})^nf^{+}f^{-}.
\end{array}
\label{e2}
\end{equation}

A one parameter deformation of the superVirasoro algebra is given by $[33]$:

\begin{equation}
\begin{array}{c}
Q^{\mathrm{l-k}}L_lL_k-Q^{\mathrm{k-l}}L_kL_l=[l-k]_QL_{k+l} \\
F_mG_n-G_nF_m=G_{n+m} \\
\,\;\;\;\;\;\;\;L_lF_k-Q^{2k}F_kL_l=-Q[[k]]_QF_{k+l} \\
Q^{\mathrm{n-m}}F_mF_n-Q^{\mathrm{m-n}}F_nF_m=\lambda [n-m]_QF_{n+m} \\
Q^{\mathrm{l-k}}L_lG_k-Q^{\mathrm{k-l}}G_kL_l=[l-k]_QG_{k+l} \\
~G_mG_n+G_nG_m=0, \\
\end{array}
\label{e5}
\end{equation}
where $[x]_Q=\frac{Q^x-Q^{-x}}{Q-Q^{-1}},\;\;\;\;[[x]]_Q=\frac{1-Q^{2x}}{%
1-Q^2}$ and $\lambda =Q-Q^{-1}$.

A possible realization of the quantum superVirasoro algebra is given as
follows:

\begin{equation}
\begin{array}{c}
L_n=-Q^{(1+\frac n2N)}(B^{+})^{n+1}B^{-} \\
G_n=Q^{(\frac n2)N}(B^{+})^{n+1}f^{+}B^{-} \\
F_n=Q^{(\frac n2)N}(B^{+})^nf^{+}f^{-},
\end{array}
\label{e7}
\end{equation}
where $B^{+}$ and $B^{-}$ the $Q$-deformed bosonic creation and bosonic
annihilation operators respectively; $f^{+}$and $f^{-}$ are the classical
fermionic ones.

Due to the property of Q-boson decomposition in the $Q\to q$ limit, the
algbra $\{B^{+},B^{-},N_B\}$ reproduces an ordinary boson $%
\{b^{+},b^{-},N_b\}$ and a k-fermion operator $\{A^{+},A^{-},N_A\}.$ In this
limit the Q-fermions become q-fermions which are object equivalent to
conventional fermion $\{f^{+},f^{-},N_f\}.$

The classical superVirassoro algebra $(sVir)$ is given, from the
classical boson $\{b^{-},b^{+},N_b\}$ and the classical fermion
$\{f^{-},f^{+},N_f\},$ by the relations $(55).$

From the operators $\{A^{+},A^{-},N_A,$ one construct the generators:

\begin{equation}
J_n=q^{-N_A}(A^{+})^{n+1}A^{-},
\end{equation}

which generates the q-deformed Virasoro algebra $(vir_q):$

\begin{equation}
\lbrack J_n,J_m]_{(q^{m-n},q^{n-m})}=[m-n]_qJ_{m+n}.
\end{equation}

It is to verify that the svir and vir$_q$ are mutually commutative:

\begin{equation}
\lbrack J_n,L_n]=[J_n,F_n]=[J_n,G_n]=0.
\end{equation}

As a results, we have the following decomposition of the quantum
superVirasoro algebra:

\[
\lim_{Q\rightarrow q}svir=vir_q\otimes svir.
\]

\section{Conclusion}

We have presented a general method to investigate the $Q\to q=e^{\frac{%
\mathrm{2\pi i}}{\mathrm{k}}}$ limit of some $Q$-deformed infinite algebras
based on the decomposition of $Q$-bosons at this limit. We note that $Q$%
-oscillator realization is crucial in this decomposition of these algebras.
We have restricted, in this work, our attention to the quatum Virasoro
algebra and quantum superVirasoro algebra. We believe that the techniques
and formulas used here will be useful to extend this study to all $Q$%
-deformed infinite Lie algebras and superalgebras. This idea will be
developed in our forthcoming paper$[34].$

\newpage\

\section*{References}

$[1]$ V.G Drinfeld, Proc. Int. Cong. Math. (Berkley, 1986), Vol 1. 798.

\noindent$[2]$ M. Jimbo, Lett. Math. Phys. 11 (1986) 247.

\noindent$[3]$ P. Kulish and E. Sklyanin, Lecture Notes in Physics, VoL 151 %
\\(Springer,1981), p. 61.

\noindent$[4]$ L. D. Fadeev, N. Reshetikhin and L. Takhtajan, Alg. Anl. 1
(1988)129.

\noindent$[5]$P. Kulish and N. Reshetikhin, lett. Math. Phys. 18 (1989) 143.

\noindent$[6]$E. Skylyanin, Uspekhi. Math. Nauk. 40 (1985) 214.

\noindent$[7]$ M. Jimbo, Int. J. Mod. Phys. A 4 (1989) 3759.

\noindent$[8]$ L. C. Biedenharn, J. Phys. A 22 (1998) L 873.

\noindent$[9]$ A. J. Macfarlane, J. Phys. A 22 (1988) 4581.

\noindent$[10]$ R. S . Dunne, A. J. Macfarlane, J. A. de Azcarraga,and J$.$%
C. PerezBueno, Phys.Lett.B387(1996)294.

\noindent$[11]$ R. S. Dunne, A. J. Macfarlane, J. A. de Azcarraga, and J. C.
Perez Bueno, hep-th/960087.

\noindent$[12]$ R. S. Dunne, A. J. Macfarlane, J. A. de Azcarraga, and J.C.
Perez Bueno, Czech. J. P. \ Phys. 46, (1996) 1145.

\noindent$[13]$ J. A. de Azcarraga, R. S. Dunne, A. J. Macfarlane and J. C.
Perez Bueno, Czech. J. P. \ Phys. 46, (1996) 1235.

\noindent$[14]$ R. S. Dunne, A Braided interpretation of fractional
supersymmetry in higher dimension ,hep-th/9703111

\noindent$[15]$ S. Majid, Introduction to braided geometry and q-Minkowski
space,hep-th/9410241.

\noindent$[16]$ S. Majid, Foundations of quantum group theory, Camb. Univ.
Press,(1995)

\noindent$[17]$ S. Majid, Anyonic Quantum groups, in spinors, Twistors,
Clifford algebras and quantum deformations(Proc. Of 2nd Max Born Symposium,
Wroclaw, Poland,1992),Z.Oziewicz et al, eds. Kluwwer.

\noindent$[18]$ R. S. Dunne, Intrinsic anyonic spin through deformed
geometry, hep-th/9703137.\ \

\noindent$[19]$ E. Ragoucy and Sorba, Int J. Mod. Phys A. 7 (1992) 2883.

\noindent$[20]$ A. Neuveu and J. Schwarz, Nucl. Phys. B. 31. (1971) 86.

\noindent$[21]$ P. Ramond, Phys. Rev. D 3 (1971) 2415.

\noindent$[22]$ N. Aizawa, T. Kobayashi, and H-T Sato, hep-th/9706176.

\noindent$[23]$ T. Kobayashi and T. Uematus, Phys. Lett. B. 306. (1993). 27;
Phys. C. 56. (1992). 193.

\noindent$[24]$ T. Kobayashi, Z. Phys. C 60. (1993). 101.

\noindent$[25]$ M. Mansour, M. Daoud and Y. Hassouni, Phys. Lett B 10 (1995).

\noindent$[26]$ M. Mansour and E. H. Zakkari, hep-th/0303215.

\noindent$[27]$ T. L. Curtright and C. K. Zachos, Phys. Lett. B 243
(1990)237.

\noindent$[28]$ A. EL Hassouni, Y. Hassouni, E.H. Tahri and M. Zakkari Mod.
Phys. Lett A 10 (1995) 2169.

\noindent$[29]$ A. El Hassouni, Y. Hassouni, E. H. Tahri and M. Zakkari,
Mod. Phys. Lett A 11 (1996) 37; M. Mansour, Inter. Jour Theor. Phys. Vol.
37, No. 9, 2467.

\noindent$[30]$ M. Chaichian, P. Kulish and J. Lukierski, Phys. Lett. B 237
(1990) 401.

\noindent$[31]$ N. Aizawa and H. Sato, Phys. Lett. B 277 (1991) 185.

\noindent$[32]$ W. S. Chung, J. Math. Phys. 35 (1994) 2490.

\noindent$[33]$ M. Mansour, Czechoslovak Journal of Physics, Vol 5(2001), N$%
^0$9, 883.

\noindent$[34]$ M. Mansour, H. Zakkari, in preparation.

\end{document}